\documentclass[twocolumn,floatfix,superscriptaddress,longbibliography,pra]{revtex4-2}

\usepackage{amssymb,amsmath,bm}
\usepackage{graphicx}
\usepackage{epstopdf}
\usepackage{color}
\usepackage{appendix}
\usepackage[colorlinks=true,citecolor=blue,linkcolor=magenta]{hyperref}
\usepackage{float}
\usepackage[normalem]{ulem}
\usepackage{bbold}
\usepackage{enumitem}
\usepackage[normalem]{ulem}

\newcommand{\ket}[1]{\vert #1 \rangle}
\newcommand{\bra}[1]{\langle #1 \vert}

\newcommand{\nb}{\nonumber}

\let\oldsection=\section

\renewcommand{\section}[1]{\emph{#1.---}}

\begin{document}

\title{Stable and efficient differentiation of tensor network algorithms}
\author{Anna Francuz}
\email{anna.elzbieta.francuz@univie.ac.at}
\affiliation{University of Vienna, Faculty of Physics, Boltzmanngasse 5, 1090 Wien, Austria}
\author{Norbert Schuch}
\affiliation{University of Vienna, Faculty of Physics, Boltzmanngasse 5, 1090 Wien, Austria}
\affiliation{University of Vienna, Faculty of Mathematics, Oskar-Morgenstern-Platz 1, 1090 Wien, Austria}
\author{Bram Vanhecke}
\email{bram.physics.vanhecke@gmail.com}
\affiliation{University of Vienna, Faculty of Physics, Boltzmanngasse 5, 1090 Wien, Austria}


\begin{abstract}
Gradient based optimization methods are the established state-of-the-art paradigm to study strongly entangled quantum systems in two dimensions with Projected Entangled Pair States. However, the key ingredient, the gradient itself, has proven challenging to calculate accurately and reliably in the case of a corner transfer matrix (CTM)-based approach. Automatic differentiation (AD), which is the best known tool for calculating the gradient, still suffers some crucial shortcomings. Some of these are known, like the problem of excessive memory usage 
and the divergences which may arise when differentiating a singular value decomposition (SVD). Importantly, we also find that there is a fundamental inaccuracy in the currently used 
backpropagation of SVD that had not been noted before. In this paper, we describe all these problems and provide them with compact and easy to implement solutions. We analyse the impact of these changes and find that the last problem -- the use of the correct gradient -- is by far the dominant one and thus should be considered a crucial patch to any AD application that makes use of an SVD for truncation. 
\end{abstract}

\maketitle
\section{Introduction}%
Tensor Network methods are designed specifically to deal with many body systems and have already proven very useful in the studies of strongly entangled states. One of their early successes was the discovery of the density matrix renormalization group (DMRG) algorithm \cite{DMRG1,DMRG2} and an independently developed matrix product state (MPS) ansatz \cite{MPS1,MPS2} that proved strikingly effective for studying one dimensional gapped spin systems. 
The two-dimensional extension of these ideas are projected entangled pair states (PEPS) \cite{PEPS}. While they hold great promise to be just as revolutionary as MPS, as can already be seen from studies of superconductivity \cite{Corboz_Hubbard,fermions3,iPEPS_tJ,vPEPS_tJ}, topologically ordered states \cite{topoAF1,topo_Corboz} and chiral spin liquids \cite{Juraj_CSL}, they still lack the ease of use that the DMRG algorithm provided for MPS. 

A key cause for the difficulty in using PEPS is that unlike for MPS, calculation of local observables cannot be done exactly, but require some form of approximate contraction of the tensor network. There are three main contraction schemes for infinite PEPS: corner transfer matrix (CTM) \cite{Baxter_CTM_78,Baxter_Textbook_82,CTMRG}, boundary MPS methods \cite{boundaryMPS} and tensor renormalization group \cite{TRG}. 
The easiest to implement and most commonly used of these schemes is the CTM. 

Optimizing PEPS for the ground state has been done in a multitude of ways. Historically, the first methods were based on imaginary time evolution~\cite{simple_update,full_update,fastfull_update,NTU},
but this in general resulted in worse states (i.e.\ worse energies and failure to capture the relevant physics) than variational methods. 
Therefore, the focus shifted towards gradient based methods. The early gradient calculations~\cite{Corboz_PEPS_gradient,Var_iPEPS2,topo_Corboz} could only approximate the gradient, so the energy minimization would not be able to converge. 
Additionally, these gradient calculations were notoriously complicated to implement.  
An important recent improvement was achieved with the use of automatic differentiation (AD) \cite{AD_CTM,Juraj_1AD}, since it greatly simplified the implementation and gave access to an exact gradient of the approximated energy, enabling a much improved convergence of the state.

The current AD procedure as applied to PEPS however still has some issues: (\emph{i}) generic AD requires a lot of memory for storing all the intermediate objects created in the iterative CTM procedure, and differentiating the fixed point equation instead suffers from gauge fixing issues, (\emph{ii}) more importantly, the gradient of the SVD -- a necessary subroutine of the CTM algorithm -- is poorly conditioned and even undefined in the case of degeneracies \cite{AD_CTM,variPEPS}, (\emph{iii}) finally, as we show, the equations used so far to backpropagate the gradient through the SVD were lacking an important term, introducing an undesired approximation. 

In this paper, we solve these problems, while keeping the inherent simplicity of the AD approach untouched. The first problem has a known solution~\cite{AD_CTM,variPEPS} where one differentiates the fixed point equation rather than the long sequence of CTM iterations. This, however, requires a reliable gauge-fixing of the CTM output, which was hitherto not possible. We identify the key issue and two methods to practically do the gauge fixing, hence fixing the memory problem. Perhaps more interestingly, we also solve the second problem -- the instability of AD for degeneracies in the SVD spectrum -- by revealing and exploiting a hidden gauge symmetry inherently present in the CTM tensor network. We show that this problem could only be solved by moving outside the conceptual scope of AD, yet the resulting solution can be implemented in a simple, AD-compatible way. Finally, we present a derivation for the gradient of a truncated SVD which differs by one term from the one previously used: 
it takes the truncated spectrum into account, whereas the previously used equation implicitly assumed the truncated spectrum to be zero as an approximation. 
We illustrate that surprisingly enough, the dominant errors are generally not those induced by degenerate spectra 
but those due to  the approximated rather than the correct gradient for the SVD, an issue which can be solved by changing a few lines in any AD code.

\begin{figure}[t!]
    \centering
    \includegraphics[width=0.9\columnwidth]{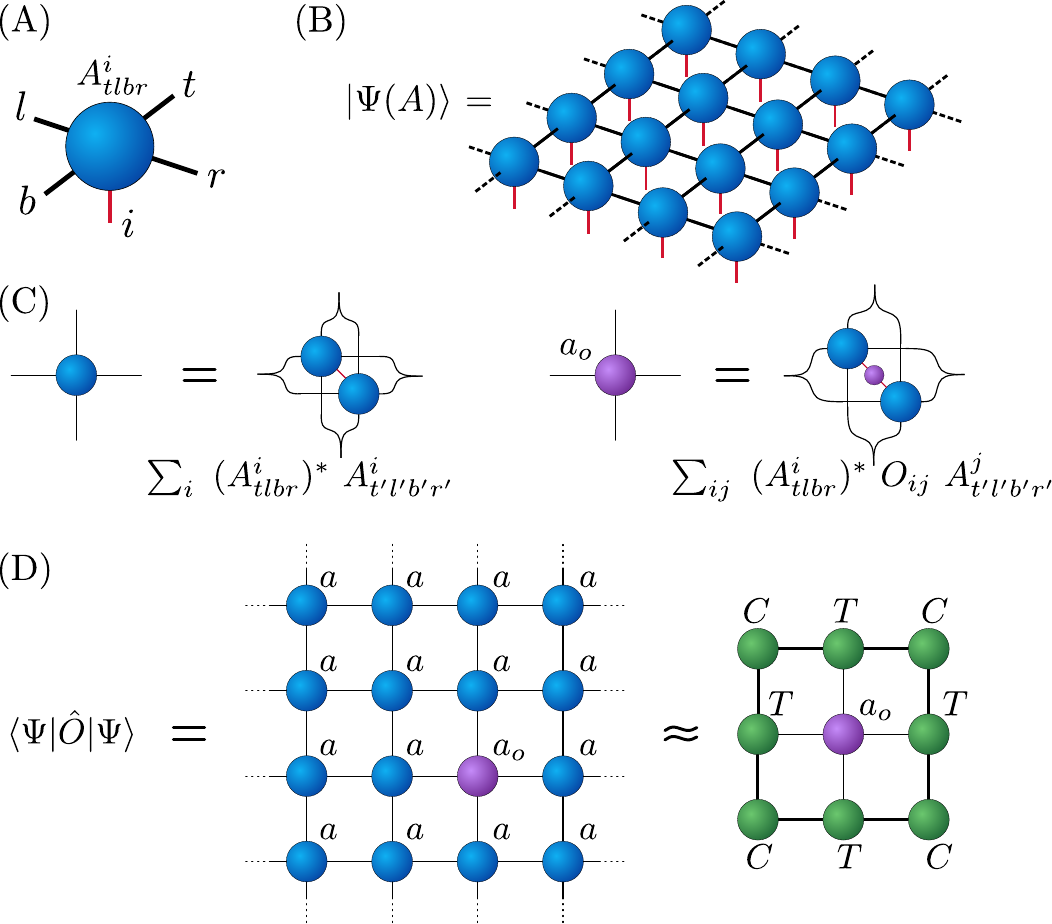}
    \caption{A state $\vert \Psi(A) \rangle$ in (B) in generated by a single PEPS tensor in (A) by contracting its virtual indices. A double PEPS tensors in (C) is formed by contracting the PEPS tensor $A$ and its conjugate via physical index and grouping the remaining open indices. In order to calculate a local observable in (D) one needs to approximate an infinite double-layer PEPS tensors network around the PEPS tensors where the operator acts with a finite environment formed by corner tensors $C$ and the edge tensors $T$.}
    \label{fig:iPEPS}
\end{figure}

\section{Symmetric CTM}%
\begin{figure}[t!]
    \centering
    \includegraphics[width=0.9\columnwidth]{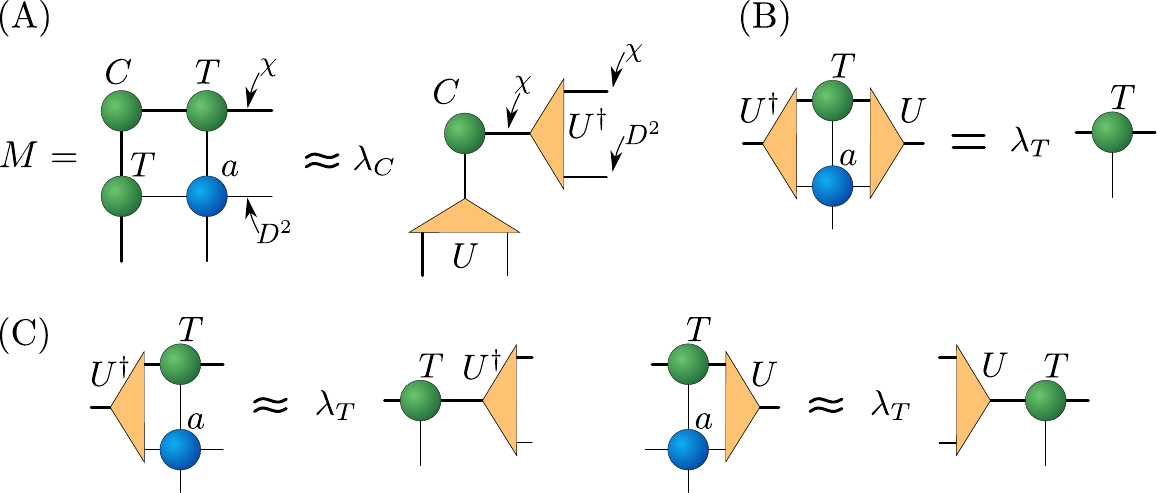}
    \caption{A single CTM update step consists of (A) diagonalization of the enlarged corner yielding a new truncated corner $C$ and new isometries $U$, which are then used to update the edge tensor $T$ in (B). In (C) the resulting renormalization identities are shown.}
    \label{fig:ctm}
\end{figure}

We will focus in this paper on a particularly simple version of CTM that relies on spatial symmetries. However, our findings have strong direct implications for all other CTM implementations.
The basic objects to calculate and optimize with PEPS are local observables $O=\bra{\Psi}\hat{O}\ket{\Psi}$, with an operator $\hat{O}$ having some finite support on few lattice sites, e.g. Hamiltonians with nearest-neighbour interactions. Without loss of generality, we focus on single site observables for now. A quantum state $\vert\Psi(A)\rangle$ encoded with infinite PEPS on a square lattice with translational invariance (Fig.~\ref{fig:iPEPS}(B)) is defined by a single site rank-5 tensor $A^i_{t,r,b,l}$ (which we assume to have C4V symmetry) with one physical index and four virtual indices (Fig.~\ref{fig:iPEPS}(A)), which are summed over in the network as depicted by the connected black lines in Fig.~\ref{fig:iPEPS}(B). A local expectation value thus depends on a tensors $A$, and can be graphically represented as in Fig.~\ref{fig:iPEPS}(C-D). We approximate this infinite tensor network using CTM. The premise of CTM, symmetric or not,  is that the `renormalisation' (CTMRG) identities shown in Fig.~\ref{fig:ctm}(C) are approximately true, and hence one can replace the infinite number of tensors surrounding the local observable with an approximation, called an 'environment', as shown in Fig.~\ref{fig:iPEPS}(D). For larger observables one would use the same CTMRG identities to derive a different environment comprised of $C$ and $T$ tensors. To find the $C$, $T$, and $U$ tensors one simply starts from some initial guesses and iteratively updates them with their respective fixed point conditions, shown in Fig.~\ref{fig:ctm}(A-B), which define the symmetric CTM algorithm~\cite{CTMRG}. In the case of symmetric CTM, the truncation tensors $U$ are determined by the eigendecomposition of the `enlarged corner' as shown in Fig.~\ref{fig:ctm}(B). This $U$ is then used to update $C$ and $T$ as shown in \ref{fig:ctm}(A) and \ref{fig:ctm}(B), respectively. If one does not have these symmetries, a more intricate scheme is needed for $U$ that involves a singular value decomposition \cite{vPEPS_tJ}.

One iteration of the CTM algorithm may be understood as a function $f$ that depends on the PEPS tensor $A$ and maps the environment tensors $C$ and $T$, which we can together represent as $x\equiv (C,T)$, to some new $x^\prime\equiv (C',T')$: $x^\prime=f(x,A)$. This is repeated until convergence, meaning an $x$ has been found that satisfies $x=f(x,A)$. This CTM environment $x_n$, converged in $n$ steps, is ultimately used to calculate an approximation of the energy for the given Hamiltonian $H$:
\begin{equation}
    E \approx \Tilde{E} = F(x_n,A,H).
\end{equation}
In order to optimize PEPS for the ground state of $H$, the above formula for the energy has to be minimized using some gradient based method. To obtain the gradient w.r.t.\ the PEPS tensor $A$ one must also differentiate $x_n$, i.e.\ $C$ and $T$ tensors, since they implicitly depend on the PEPS tensor through the iterative formula described above, $x_n = f(x_{n-1},A)$. This intricacy of the dependence of $C$ and $T$ on the PEPS tensor is what makes PEPS optimization hard, and what makes automatic differentiation an attractive option.

\section{Automatic differentiation}%
The simplest AD framework, available in many programming languages, makes use of the chain rule applied to the approximate energy $\Tilde{E}$ calculated using converged $C$ and $T$ tensors obtained from the above described CTM procedure. It goes back through this process, i.e.\ through all the iterations of the CTM, to collect contributions to the gradient from each step: 
\begin{eqnarray}
    d\Tilde{E} &=& \frac{\partial F}{\partial A} dA + \frac{\partial F}{\partial x_n} dx_n,   \label{eq:dE} \\
    dx_n &=&\frac{\partial f}{\partial A} dA + \frac{\partial f}{\partial x_{n-1}}\left(\frac{\partial f}{\partial A}dA + \frac{\partial f}{\partial x_{n-2}} (...)\right), \label{eq:dxn}
\end{eqnarray}
where at each step we substituted $x_k$ by $f(x_{k-1},A)$. While this may be algorithmically complicated, it can be done completely automatically (black box). However, using the black box leads to two known issues. Firstly, instead of initializing the CTM algorithm with a good initial guess (e.g. $C,T$ from a previous iteration in the gradient optimization), which generally leads to faster convergence, it is necessary to start every time with random initial $C,T$ to guarantee enough iterations and no dependence on the initial conditions. Secondly, it requires storing all intermediate objects, which can be extremely memory intensive. The solution to these problems is to differentiate the fixed point condition $x=f(x,A)$ instead \cite{AD_CTM,ImplicitFunction}. This may in principle be done with the same practical ease as differentiating through the iterations using the tools of AD. However, this requires gauge fixing of the CTM result, which will be discussed in the next section.

\section{Fixed point differentiation}%
As was previously realised~\cite{AD_CTM,ImplicitFunction}, one can differentiate the fixed point equation for $x = f(x,A)$, rather than differentiating all the iterations that led to that solution. The derivative can then be written as a geometric series:
\begin{equation}
    dx = \sum\limits_{k=0}^\infty \left(\frac{\partial f}{\partial x} \right)^k \frac{\partial f}{\partial A}dA = \left(1-\frac{\partial f}{\partial x} \right)^{-1} \frac{\partial f}{\partial A}dA.
    \label{eq:fp_diff}
\end{equation}
The operation of $\frac{\partial f}{\partial x}$ can be implemented using AD.
An important aspect of fixed point differentiation is that $x$, i.e. $C$ and $T$, must have converged element-wise, which does not generally happen with CTM. Instead, CTM converges to an $x=f(x',A)$ where $x'$ is related to $x$ by a gauge transform.
The origin of this arbitrary freedom is the non-uniqueness of the eigenvalue decomposition. Indeed, we may transform any eigenbasis $U$ as follows:
\begin{equation}
\label{eq:eigenval-gauge-dof}
UCU^\dagger = \big(U\sigma\big) C\big(\sigma^\dagger U^\dagger\big),
\end{equation}
where $\sigma$ is any unitary matrix satisfying $\left[C,\sigma\right]=0$. Thus $C$ converges element-wise (if the eigenvalues are uniquely sorted) but $T$ can differ from iteration to iteration by a random gauge $\sigma$:
\begin{equation}
\hat{T} \stackrel{f}{\to}
\sigma^\dagger \hat{T} \sigma=T,
\label{eq:Tgauge}
\end{equation}
where $\hat{T}$ is an edge tensor from the previous CTM iteration. It is tempting to think that since the gauge freedom of the eigendecomposition,
Eq.~\eqref{eq:eigenval-gauge-dof},
is the origin of the problem, it can be resolved by making the eigendecomposition unique in some way. However, for differentiating the fixed point condition, it is rather relevant that $T$ (and $C$) are fixed points, and not the eigenvalue decomposition itself. 
Thus, one should rather determine a gauge such that the last CTM step maps $\hat{T}$ to $T$. In practice, one takes the last CTM step with some arbitrary gauge and deduce a $\sigma$ that relates the two $T$ tensors as in Eq.~\eqref{eq:Tgauge}. Incorporating it into the isometry, i.e.,  $U~\rightarrow U\sigma$, we get a gauge-fixed eigendecomposition that leads to the desired element-wise convergence of $T$ so Eq.~\eqref{eq:fp_diff} can be utilised and the memory problem solved. 
Details of the gauge fixing schemes used in this work are presented in Appendix \ref{app:gauge_fixing}, and apply to any CTM. 

\begin{figure}[t!]
    \centering
    \includegraphics[width=\columnwidth]{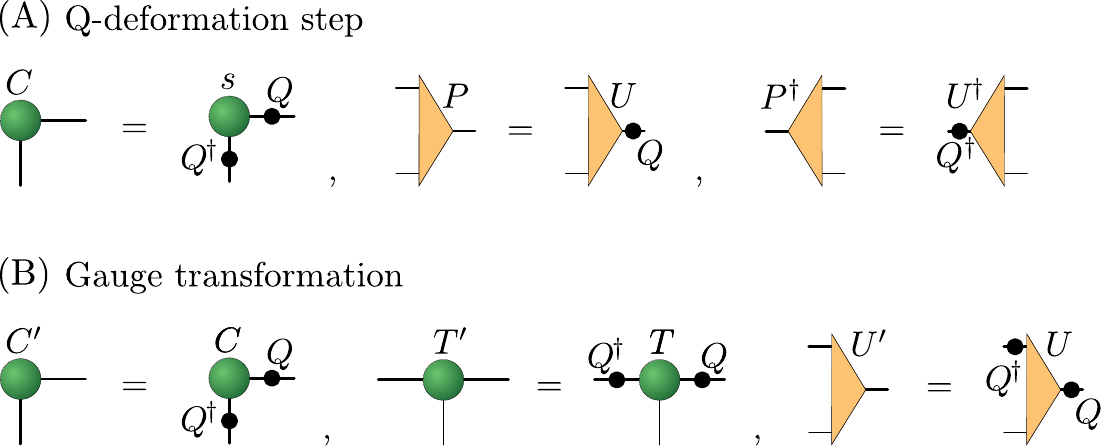}
    \caption{Q-deformation in (A) results in a CTM fixed point related to the original one by a gauge transformation in (B).}
    \label{fig:deformation}
\end{figure}

\section{Q-deformed CTM}\label{free_gauge}%
The second issue with current AD implementations which we tackle is that the gradient of the eigendecomposition contains a diverging term in case of a degenerate spectrum. It is currently treated by slight deformation of the spectrum through e.g.\ Lorentzian broadening~\cite{AD_CTM}. However, the ill-conditioned term is clearly an artefact of the method: Given a PEPS for which the $C$ matrix has a degeneracy, it should be possible to perturb the PEPS tensor to slightly break this degeneracy, while at the same time smoothly changing the energy $\Tilde{E}$, thereby effectively removing the divergence. This strongly suggests that  a better method exists which avoids these divergences altogether.

In order to devise such a divergence-free method, we make use of a 
hitherto ignored gauge degree of freedom in the CTM iteration. Specifically, 
in the truncation of the enlarged corner $M$, Fig.~\ref{fig:ctm}(A), the current CTM implementations choose the isometries $U$ such that the new corner $U^\dagger M U=C\equiv s$ is diagonal,  while in principle, any isometry projecting onto the subspace with the $\chi$ largest eigenvalues would be fine. We can re-introduce this basis choice degree of freedom through a unitary $Q$, by defining a new isometry $P=UQ$. Since $Q$ is simply a basis change, 
the new fixed point tensors of the Q-deformed CTM (i.e., under this change of isometry), 
 $C'$ and $T'$, are related to the fixed points $C$ and $T$ 
 of the original CTM simply by conjugating with $Q$, 
Fig.~\ref{fig:deformation}(B); specifically, $C'=Q^\dagger s Q$. It is thus clear that the $Q$-deformed CTM algorithm will converge 
to the same, albeit gauge transformed, result, and in particular keeps the observables unchanged. 

In the following, we will use this additional degree of freedom $Q$ to remove the divergences which appear in the differentiation of normal (undeformed) CTM. To this end, we will set $Q=\openone$, but make use of the \emph{variations} 
$dQ$ of $Q$ (which don't affect any physical property, since $Q$ is a gauge degree of freedom) to cancel the divergent terms. (This also shows that the ad-hoc gauge fixing for the isometries $U$ is the source of the divergence, and that one must generally be rather careful when fixing gauges without reason.)

\section{Stable derivative}%
To differentiate the $Q$-deformed CTM step, i.e.\ calculate $\frac{\partial f}{\partial x}$ and $\frac{\partial f}{\partial A}$ introduced earlier, one mainly needs to differentiate tensor contractions. These can be easily handled with AD tools, and are not affected by the insertion of $QQ^\dagger=\openone$. 
The one step which needs special attention to avoid divergences is the eigendecomposition and truncation of the enlarged corner $M$, Fig.~\ref{fig:ctm}(A), which we discuss in the following.

We start the derivation with the full decomposition of the enlarged corner $M$ of the $Q$-deformed CTM, shown in Fig.~\ref{fig:ctm}(A):
\begin{equation}
    M=P C P^\dagger + P_\perp C_\perp P_\perp^\dagger\ ,
    \label{eq:Mdecomp}
\end{equation}
where $P = UQ$ projects onto the leading eigenvalue subspace, in a basis where the corner is not diagonal, i.e.\ $C=Q^\dagger s Q$ with $s$ the diagonal matrix with the dominant eigenvalues, 
and $P_\perp C_\perp P_\perp^\dagger$ representing the truncated part of the eigendecomposition. The latter does not have to be calculated explicitly but is introduced here for the analysis. Note that the truncated eigen-/singular values were never taken into account in previous applications of AD to CTM~\cite{AD_CTM,Juraj_1AD,variPEPS}, hence our result will differ regardless of the inclusion of $Q$.

Given that $Q$ is unitary, $U$, $P$ and $P_\perp$ are isometries, and $P^\dagger P_\perp=0$, we can parameterise their infinitesimal variations as follows:
\begin{align}
dQ&=d\omega' Q \\
dU&=U_\perp dX + U d\omega,
\label{eq:Uparametrization}
\end{align}
and thus
\begin{equation}
dP=d(UQ) = U_\perp dX Q + U(d\omega +d\omega')Q 
     \label{eq:Pparametrization}
\end{equation}
with $d\omega$, $d\omega'$ anti-Hermitian generators of rotations within the subspace defined by $P$, and $U_\perp dX$ parameterising an infinitesimal rotation into the subspace orthogonal to the one spanned by $P$. The divergences we are aiming to remove appear in $d\omega$; for completeness, we provide the explicit form in Appendix~\ref{appendix:differentiate-eigen}. 
However, we can now simply cancel $d\omega$ in 
Eq.~\eqref{eq:Pparametrization} by choosing $d\omega'=-d\omega$, i.e., through a suitable variation of $Q$. Importantly, 
any choice for the gauge freedom $Q$ and $\frac{\partial Q}{\partial A}$ is legitimate, as it does not affect the energy $E$.  

We can then calculate the variation $dC$, which is not diagonal, by taking the derivative of Eq.~\eqref{eq:Mdecomp} and projecting onto $P^\dagger$ from the left and $P$ from the right:
\begin{equation}
    P^\dagger dM P=dC\ .
    \label{eq:dC}
\end{equation}
The variation of $P$ follows from applying $U_\perp^\dagger$ and $P$ to the left and right of $dM$, respectively, and using the fact that $dU^\dagger_\perp P= - U^\dagger_\perp dP$ (this follows from differentiating $U^\dagger_\perp P = 0$):
\begin{align}
    U_\perp^\dagger dM P&=dX C + s_\perp dU^\dagger_\perp P\\
    \Rightarrow\big(\mathbb{1}-PP^\dagger\big)dMP&=dPC-\underline{\big(\mathbb{1}-PP^\dagger\big)MdP}
    \label{eq:dP}
\end{align}
This Sylvester equation (an equation for a matrix $Y$ of the form $A Y + YB = D$) for $dP$ has a unique solution if $C$ and $C_\perp$ have strictly different spectra. One should thus be careful not to split multiplets when converging the original CTM.

Remarkably, one never needs to calculate either $d\omega'$ or $d\omega$. Nor, in fact, is a non-trivial $Q$ needed, as it does not appear in any of the final equations (\ref{eq:dC},\ref{eq:dP}). So one can effectively use a regular CTM algorithm, which is a $Q$-deformed CTM with $Q=\mathbb{1}$, and still make use of a non-trivial $\frac{\partial Q}{\partial A}$. Our result can thus be summarised as providing a modified formula for the gradient of the eigendecomposition with a $dC$ that is non-diagonal, where all the instabilities and divergences are removed. All the rest of the CTM differentiation can remain unchanged. We give some practical details in the appendix \ref{app:practical_gradient} about how to efficiently use Eq.~\eqref{eq:dP} in backwards differentiation.

Let us emphasize that our result represents an improvement over the previously used AD in two independent ways: First, the $Q$-deformation makes the rightmost (divergent) term in Eq.~\eqref{eq:Pparametrization} vanish, and second, the inclusion of the truncated part of $M$ in Eq.~\eqref{eq:Mdecomp} gives rise to the additional underlined term in Eq.~\eqref{eq:dP}.
The effects  of cancelling the divergent $d\omega$ and of including the extra term  will be benchmarked independently later.

\section{Gradient of SVD}%
Using the same techniques we used above to differentiate a truncated eigendecomposition, we can also differentiate the SVD. Here, we just give the part (underlined) of the result that differs from the one previously used. A complete derivation can be found in Appendix~\ref{appendix:differentiate-svd}.
\begin{align}
    (\mathbb{1}&-U U^\dagger) dAV=(\mathbb{1}-U U^\dagger)(dUS - \underline{AdV})\\
    (\mathbb{1}&-V V^\dagger) dA^\dagger U=(\mathbb{1}-V V^\dagger)(dVS - \underline{A^\dagger dU})\label{eq:svd_grad2}
\end{align}
Both underlined terms on the right-hand side were not included in previous implementations.

\begin{figure}[t!]
    \centering
    \includegraphics[width=0.95\columnwidth]{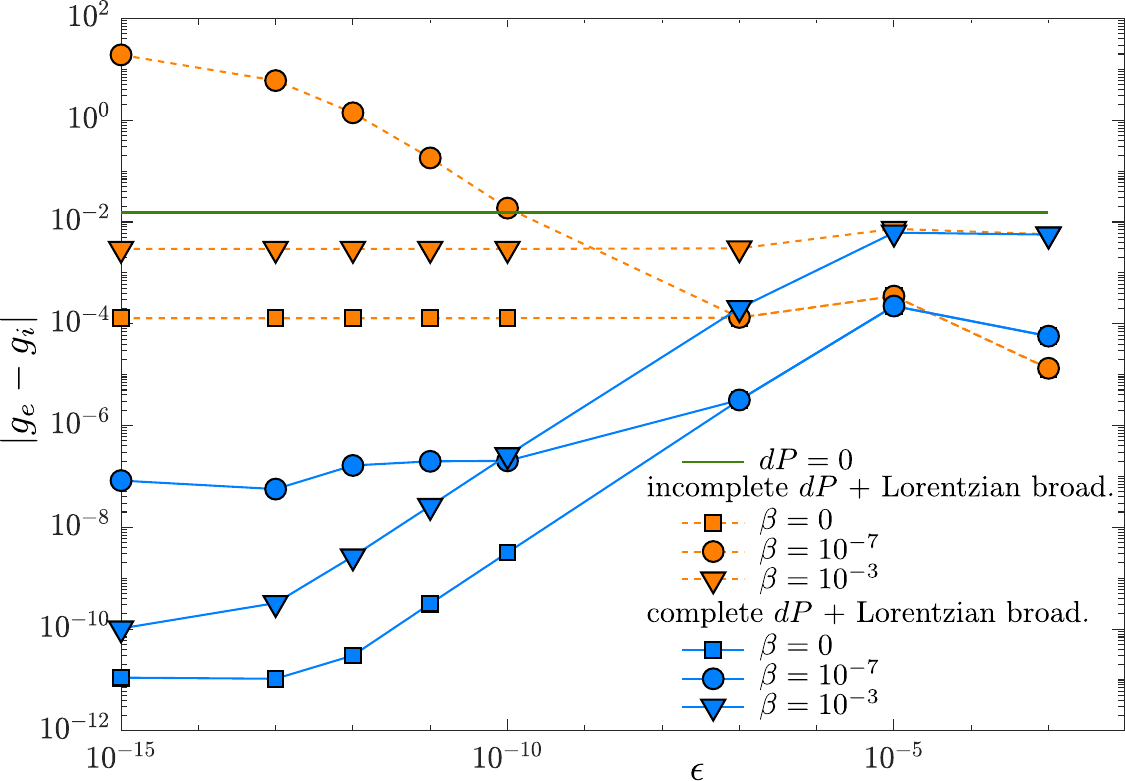}
    \caption{Difference between the exact gradient $g_e$ obtained using all techniques derived in this work, and incomplete gradients $g_i$ obtained
    using various approximations (see text), for different degeneracy-lifting perturbations $\beta$ and Lorentzian broadenings $\epsilon$. By far the largest improvement can be achieved by including the complete $dP$ in the gradient (blue lines).}
    \label{fig:gradients}
\end{figure}


\section{Benchmark}
Let us now benchmark the improvement in the gradient gained through our findings, compared to existing approaches. To this end, we pick a Hamiltonian and a (not optimal) PEPS ansatz which forms an initial or intermediate state for a gradient optimization in CTM. We then compute the CTM fixed point using our gauge fixing scheme (from Appendix \ref{app:gauge_fixing_robust}), and differentiate it to compute the gradient using different approaches, covering existing approaches as well as some or all of our improvements. 
To be able to study the effect of (approximate) degeneracies in the spectrum of $C$, we choose an SU(2)-symmetric ansatz with long correlation length (leading to truncation at relatively large eigenvalues), to which we add a symmetry-breaking perturbation in order to induce a tunable splitting of the multiplets.

Specifically, the ansatz we choose is a  one-parameter family of perturbed RVB states with nearest neighbour and longer range singlets with C4V symmetry, which can be expressed as PEPS with bond dimension $D=3$~\cite{rvb}:
\begin{equation}
    \mathcal{A} = \mathcal{A}_1 +\mathcal{A}_2 + \beta \mathcal A_3\
\end{equation}
where the $SU(2)$ symmetric tensors $\mathcal{A}_1,\mathcal{A}_2$ generate the NN RVB state and longer range singlets, respectively, and $\beta$ controls an SU(2) symmetry breaking perturbation $\mathcal A_3$. We compute the gradient with respect to the Heisenberg $J_1-J_2$ Hamiltonian on a square lattice with single site unit cell:
\begin{equation}
    H = \sum\limits_{i,j\in \mathrm{NN}, \alpha} f(\alpha) S^\alpha_iS^\alpha_j + \frac{J_2}{J_1} \sum\limits_{i,j\in \mathrm{NNN}} \vec{S}_i\cdot\vec{S}_j 
\end{equation}
with $SU(2)$ symmetry breaking anisotropy $f([x,y,z])=[-1,1+\beta,-1+\beta]$ controlled by the same parameter $\beta$, and $J_2/J_1 = 0.5$. We calculate the CTM environment with bond dimension $\chi=160$. Our gradient calculations agree with the gradient obtained from $4^\text{th}$ order finite difference calculations up to $\mathcal{O}(10^{-12})$, which verifies that it is exact. In Fig.~\ref{fig:gradients}, we show how much the following approximate ways to determine the gradient deviate from our exact and divergence-free gradient: 
\begin{itemize}[leftmargin=*]
    \item a gradient assuming $dP=0$, as done in \cite{topo_Corboz} before the AD framework was introduced (green line)
    \item the currently available AD framework with incomplete gradient of $dP$, using Lorentzian broadening $\frac{1}{x_{ij}}\to\frac{x_{ij}}{x^2_{ij}-\epsilon}$ with $x_{ij} = s_j-s_i$ to 
    avoid divergences \cite{AD_CTM,Juraj_1AD,variPEPS} (orange lines)
    \item a gradient based on the complete differential of $dP$, obtained from Eq.~\eqref{eq:dP} (i.e.\ including the discarded part of the spectrum in Eq.~\eqref{eq:Mdecomp}), but still using Lorentzian broadening to avoid divergences (blue lines)
\end{itemize}
all of which were calculated using the fixed point differentiation we described earlier.

Fig.~\ref{fig:gradients} shows the difference to the exact and divergence-free gradient as a function of the parameter $\epsilon$ of the Lorentzian broadening for three different values of the SU(2)-breaking perturbation $\beta=0,10^{-7},10^{-3}$. Remarkably, we find that by far the strongest impact comes from whether we use the full expression for $dP$, i.e., the correct gradient, or rather the previously used approximation for $dP$, as can be seen from the clear separation of the blue vs.\ orange lines in Fig.~\ref{fig:gradients}. 
In fact, we noticed that the linear problem defined by Eq.~\eqref{eq:fp_diff} for the fixed point differentiation can become ill-conditioned if the full equation for $dP$ is not used. 
Beyond that, we also observe that in the case of approximate degeneracies in the spectrum ($\beta=10^{-7}$), using our singularity-free truncation performs significantly better, while both for exact degeneracies ($\beta=0$) or for strongly broken degeneracies ($\beta=10^{-3}$), the improvement gained over Lorentzian broadening is minor.

\section{Conclusions}%
We have studied two central problems in the application of AD to tensor network algorithms, in the course of which we identified a third key problem, and devised solutions to all three of them. 
While we focused on CTM, our findings apply to AD of any algorithm which involves the truncation of eigenvalue or singular value decompositions. We have analyzed the effect of these solutions on the  performance of the CTM algorithm, and observed that by far the largest effect is achieved by solving the third problem, that is, the use of the correct gradient. This can be achieved by adding just a few lines of code, and should thus be considered an essential fix to any AD-based tensor network algorithm.

\section{Acknowledgements}%
The authors are grateful to Boris Ponsioen, Juraj Hasik, Paul Brehmer and Andras Molnar for stimulating discussions. 
Numerical calculations were performed in MATLAB with the help of the \verb+ncon+ function \cite{ncon} for tensor contractions. 
This research was funded in part by the Austrian Science Fund (FWF) [projects \mbox{ESP-306}, P-36305, F-71], and the European Union’s Horizon 2020 program through Grant No.\ 863476 (ERC-CoG \mbox{SEQUAM}).

\let\section=\oldsection

\clearpage
\section*{Appendices}
\appendix

\section{Gauge fixing -  Generic} 
\label{app:gauge_fixing}

In the following two appendices, we discuss methods to properly gauge fix the fixed point of the CTM algorithm, to utilize the element-wise convergence and the fixed point differentiation from Eq.~\eqref{eq:fp_diff}. The output of the CTM is already partially gauge-fixed by setting $C$ to be a diagonal matrix (but still allowing for $dC$ to be non-diagonal), so we are left with the remaining gauge freedom $\sigma$ from Eq.~\eqref{eq:Tgauge} due to the phase freedom of the eigenvectors, which we recall here for clarity:
\begin{equation}
    T = \sigma^\dagger\hat{T}\sigma,
\end{equation}
and where $\hat{T}$ is an edge tensor from the previous iteration. When the corner $C$ is non-degenerate, the problem of determining $\sigma$ simplifies to finding a diagonal matrix of phases, in which case the equation simplifies to (with $i$ being the physical index):
\begin{equation}
\forall_i ~ \hat{T}^i_{ab} = \sigma_{aa} T^i_{ab} \sigma^*_{bb}\ .
\end{equation}
Without loss of generality, we set $\sigma_{11} = 1$ and solve for $\sigma_{aa},~ a\neq 1$:
\begin{equation}
\sigma_{aa} = \text{sgn}\left(\frac{\hat{T}^i_{1a}}{T^i_{1a}}\right)\ .
\end{equation}
If the tensor $T$ happens to be sparse (i.e. $T_{1a} = 0$ for some $a$), it may just take more iterations to find all the diagonal elements of $\sigma$ using those previously obtained from above equation:
\begin{equation}
\sigma_{bb} = \text{sgn}\left(\frac{\sigma_{aa}\hat{T}^i_{ab}}{T^i_{ab}}\right).
\end{equation}

However, this approach is not general enough and the problem becomes much more complicated if the spectrum of the corner tensor $C$ is degenerate, in which case $\sigma$ is a block diagonal matrix of continuous rotations within the degenerate subspaces. Therefore, we proceed by examining the eigenvectors (or singular vectors if the transfer matrix is non-diagonalizable) of the transfer matrix for the overlap of $T$ and $\hat{T}$ with a random MPS $M$ (which should thus be injective):
\begin{equation}
\mathbb{T} = \sum_i T^i \otimes (M^i)^*, \qquad \hat{\mathbb{T}} = \sum_i \hat{T}^i \otimes (M^i)^*\ .
\end{equation}
Choosing a random MPS can help in removing possible degeneracies in the spectrum of the transfer matrices. Note that the bond dimension of $M$ should be at least as big as that of $T$. If the transfer matrices are non-degenerate and its leading eigenvectors $\rho$  and $\hat{\rho}$ are full rank and unique then they are related by:
\begin{equation}
\hat{\rho} = \sigma\cdot \rho ~\Rightarrow~ \sigma = \rho \hat{\rho}^{-1}.
\end{equation}
To prevent having to do this inverse, we propose the following: If the tensors $T,\hat{T}$ are equal up to the gauge transformation, then the upper triangular matrices from the QR decomposition of the fixed points $\rho$ and $\rho'$ must be equal, provided that the diagonal is positive. Then the equation simplifies to: 
\begin{equation}
\sigma = Q_1 R R^{-1} Q_2^\dagger = Q_1 Q_2^\dagger\ .
\end{equation}
In fact, $\rho$, $\hat{\rho}$  could be any eigenvectors to the same unique eigenvalue (or singular vectors to the same unique singular value), not only the leading ones. This can become advantageous in cases when the leading eigenvectors are not full rank or not unique.

\section{Gauge fixing -- Robust}
\label{app:gauge_fixing_robust}
The gauge fixing methods described in the previous appendix may run into trouble in special cases when the corner is degenerate and the $T$ tensor additionally has some virtual symmetries. Therefore, in this appendix, we present a bit more involved algorithm which turns out to be the most robust gauge fixing scheme in general.
In order to fully fix the gauge of the edge tensor $T$, one has to solve simultaneously the following equations:
\begin{eqnarray}
\hat{T}\sigma &=& \sigma T, \nb \\
\left[\sigma,C \right] &=& 0, \nb \\
\sigma^\dagger \sigma &=& \mathbb{1} .
\label{eq:gauge_fix_r}
\end{eqnarray}
The first equation is a linear equation, where $\sigma$ can be reinterpreted as a vector in the null space of the vectorized equation:
\begin{equation}
\label{eq:app:gauge-vectorized}
    \left(\hat{T} \otimes \mathbb{1} - \mathbb{1}\otimes T^T\right)\vec{\sigma} = 0,
\end{equation}
while the second one limits the number of elements of $\vec{\sigma}$ needed to solve for, as it assures that $\sigma$ has a block diagonal form with zeros everywhere else. Therefore, instead of decomposing a full $\chi^2 D^2 \times \chi^2$ matrix, we only need to find the null space of Eq.~\eqref{eq:app:gauge-vectorized}, which is the same as the right null space of $\left(\hat{T} \otimes \mathbb{1} - \mathbb{1}\otimes T^T\right)^\dagger\left(\hat{T} \otimes \mathbb{1} - \mathbb{1}\otimes T^T\right)$. Additionally, we can restrict consecutively to all the degenerate subspaces of the corner $C$ for finding the right null space, which together with the sparse structure allows for efficient singular value decomposition. Finally, vectors from the null space of the reduced vectorized equation form block diagonal matrices $\sigma^i$, whose linear combination $c_i\sigma^i$ forms a unitary matrix $\sigma$, which is a solution to Eq.~\eqref{eq:gauge_fix_r}:
\begin{equation}
    \sum_{\beta}\sum_{ij} (c_i\sigma^i_{\alpha,\beta})^\dagger (c_j\sigma^j_{\beta,\gamma}) = \delta_{\alpha,\gamma}.
    \label{eq:unitarity}
\end{equation}
In case that the $T$ tensor has some virtual symmetries of the form $T = xTx^{-1}$, the null space  necessarily has dimension larger than 1 and Eq.~\eqref{eq:gauge_fix_r} may have more than one solution, but we only need to find one of them for the gauge fixing scheme and calculation of the derivative. In order to do that, we use the Levenberg-Marquardt algorithm to find coefficients $c_i$ from Eq.~\eqref{eq:unitarity}.

A note on gauge fixing for general CTM: For more general CTM applicable for larger unit cells, we may trivially generalise the first `generic' gauge fixing method by taking a row/column of edge tensors as the unit cell of an infinite MPS. The robust approach may similarly be generalised by replacing the first equation in Eq.~\eqref{eq:gauge_fix_r} with $\hat{T}_1...\hat{T}_n\sigma = \sigma T_1...T_n$. Using the aforementioned tricks, the null space problem remains of the same size as before.

\section{Differentiating a truncated eigendecomposition}
\label{appendix:differentiate-eigen}
In this appendix, we expand on the derivation of the stable gradient of eigendecomposition step in the CTM algorithm making use of the gauge freedom introduced in the $Q$-deformed CTM, and relate it to the previous approaches. The notation follows that in the main body of the paper.

There are two main points that add to the stability and accuracy of the gradient calculation. Firstly, by introducing a $Q$-deformed CTM, we realise that one may still keep $C$ diagonal; yet, by allowing its derivative $dC$ to be non-diagonal we can get rid of a problematic, possibly divergent term which we are going to derive later. This is achieved by introducing gauge rotations $Q$ within the relevant subspace of the enlarged corner eigendecomposition:
\begin{equation}
    UsU^\dagger = UQQ^\dagger s QQ^\dagger U^\dagger = PCP^\dagger.
\end{equation}
The parametrization of $dU$ (and hence $dP$) is then obtained by multiplying it with a resolution of the identity:
\begin{eqnarray}
dU &=& (UU^\dagger + U_\perp U_\perp^\dagger) dU = Ud\omega + U_\perp dX.\\
    dP &=& d(UQ) = U_\perp dX Q + U(d\omega+d\omega')Q, \label{eq:dP_parametrization}
\end{eqnarray}
where $d\omega$ is the problematic term. Introducing gauge rotations $Q$ and giving them an arbitrary dependence on the tensors A, such that $d\omega'\propto \frac{\partial Q}{\partial A}$, we use it to fully cancel the second term in Eq.~\eqref{eq:dP_parametrization} (this is possible since $\frac{\partial \Tilde{E}}{\partial Q} = 0$, i.e., the energy does not depend on $Q$). Moreover, we also know that $\Tilde{E}(Q,A) = \Tilde{E}(\mathbb{1},A)$, which allows us to set $Q=\mathbb{1}$.

Secondly, we perform our derivation using the spectral decomposition of the enlarged corner without assuming that the leading eigenspace is already a good enough approximation by itself (which it is clearly not for systems close to or at the criticality). Therefore, the full eigendecomposition of the enlarged corner $M$ can be written as in Eq.~\eqref{eq:Mdecomp}:
\begin{equation}
\label{eq:app:diff-eigen-M}
    M = PCP^\dagger + P_\perp s_\perp P_\perp^\dagger.
\end{equation}
Here, we assume from the beginning that $C$ and $dC$ do not have to be diagonal -- it only matters that the isometries $P$ and $P_\perp$ map onto relevant and truncated subspaces, respectively. Here, the square matrix $V=(P,P_\perp)$ is unitary, $V^\dagger V = V V^\dagger = \mathbb{1}$. We differentiate Eq.~\eqref{eq:app:diff-eigen-M} to obtain:
\begin{eqnarray}
    dM &=& dPCP^\dagger + PdCP^\dagger + PCdP^\dagger + \nonumber \\
        && dP_\perp s_\perp P_\perp^\dagger + P_\perp ds_\perp P_\perp^\dagger + P_\perp s_\perp dP_\perp^\dagger\ .
        \label{eq:dM}
\end{eqnarray}
 Since we moved all the rotational dependence from $dP$ to $dC$, the only contributing perturbations are those which are mixing between the relevant and truncated subspaces, i.e.,  $dP = P_\perp dX$.
By projecting Eq.~\eqref{eq:dM} onto $P^\dagger$ from the left and $P$ from the right, we obtain the non-diagonal $dC$ derivative as in Eq.~\eqref{eq:dC}:
\begin{equation}
    P^\dagger dM P = dC.
\end{equation}
Note that by expanding the right-hand side of this equation as $dC = d(Q^\dagger s Q)$, we would arrive back to the familiar formula for the derivative of the eigenvalue decomposition:
\begin{equation}
    (PQ^\dagger)^\dagger dM (PQ^\dagger) = -d\omega' s + ds + sd\omega'.
\end{equation}
Dividing this equation into its diagonal and off-diagonal part results precisely in the same formulas which are known to the AD community \cite{Matrix_derivatives}, with $d\omega=-d\omega'$:
\begin{eqnarray}
ds &=& \mathbb{1}\circ\left( U^\dagger dM U\right) \\
d\omega' &=& F^T\circ \left( U^\dagger dM U\right), \quad F_{ij} = \frac{1}{s_j-s_i},
\end{eqnarray}
(and $F_{ii}=0$) 
where the equation for $d\omega'$ is clearly divergent if the spectrum of the enlarged corner is degenerate. 
(Here, $\circ$ denotes the Hadamard product, i.e., the element-wise multiplication of matrices.)
In any of the approaches, the $P_\perp dX$ term can be obtained by projecting $dM$ onto $P_\perp P_\perp^\dagger$ from the left and $P$ from the right:
\begin{eqnarray}
    P_\perp P_\perp^\dagger dM P &=& P_\perp dX C - P_\perp s_\perp dX, \nonumber \\
    (\mathbb{1} - PP^\dagger) dM P &=& dP C - P_\perp s_\perp P_\perp^\dagger P_\perp dX \nonumber\\
     (\mathbb{1} - PP^\dagger) dM P &=&dP C - (\mathbb{1} - PP^\dagger)M dP,
     \label{eq:app_dP}
\end{eqnarray}
which results in a Sylvester equation for $dP$. In practice, there is no need for explicit construction of the enlarged corner $M$. In the end, we only need to solve a Sylvester equation numerically, knowing how the enlarged corner matrix acts on the proper adjoint, so that the backward differentiation process has the same computational cost as the forward CTM algorithm.

\section{Differentiating a truncated singular value decomposition}
\label{appendix:differentiate-svd}

In the main text, we have observed that the dominant error in the gradient comes from not including the truncated part of the eigenspectrum in the derivation. This also has direct consequences for CTM schemes that do not assume C4V symmetry of the PEPS and thus use a singular value decomposition (SVD) instead.  In this appendix, we derive an exact gradient of the truncated SVD. 

We starting from the following equation
\begin{equation}
    A=USV^\dagger + U_\perp S_\perp V^\dagger_\perp\ .
\end{equation}
Since $U$, $V$, $U_\perp$ and $V_\perp$ are all isometries we have that $U^\dagger dU$, $V^\dagger dV$, $U_\perp^\dagger dU_\perp$, and $V_\perp^\dagger dV_\perp$ are all anti-Hermitian. Additionally, we use that the orthogonality conditions must all remain satisfied, so $dU_\perp^\dagger U + U_\perp^\dagger dU=0$ and $dV_\perp^\dagger V + V_\perp^\dagger dV=0$.

It is convenient to split the parts of $dU$ and $dV$ as:
\begin{align}
    dU&=(\mathbb{1}-U U^\dagger)dU + U U^\dagger dU\\
    &=dU_2 + U dU_1\\
    dV&=(\mathbb{1}-V V^\dagger)dV + V V^\dagger dV\\
    &=dV_2 + V dV_1\label{eq:dV2}\ .
\end{align}
We start by applying $(\mathbb{1}-UU^\dagger)$ and $V$ to $dA$:
\begin{align}
    (\mathbb{1}-U U^\dagger) dAV&=(\mathbb{1}-U U^\dagger) dUS +  U_\perp S_\perp dV^\dagger_\perp V\\
    &=dU_2S - (\mathbb{1}-U U^\dagger)A\,dV_2\label{eq:dU2}\ .
\end{align}
Next we apply $(\mathbb{1}-V V^\dagger)$ and $U$ to $dA^\dagger$:
\begin{align}
    (\mathbb{1}-V V^\dagger) dA^\dagger U&=(\mathbb{1}-V V^\dagger) dVS +  V_\perp S_\perp dU^\dagger_\perp U\\
    &=dV_2S - (\mathbb{1}-V V^\dagger)A^\dagger dU_2\ .
\end{align}
These two equations allow us to solve for $dU_2$ and $dV_2$. 
Next, we apply $U^\dagger$ and $V$ to $dA$:
\begin{align}
    U^\dagger dAV&=U^\dagger dUS +  dS + SdV^\dagger V \\
    &=dU_1 S + dS + S dV_1^\dagger\ .
    \label{eq:UdAV}
\end{align}

We pause here to point out the gauge freedom inherent in the singular value decomposition, noted before in \cite{complexSVD}. The relative phases of the singular vectors in $U$ and $V$ are in principle free. This implies that the change in phase of the singular vectors of $U$, reflected in the diagonal part of $dU_1$, may be varied arbitrarily if the corresponding phases of the singular vectors in $V$ are varied accordingly. We can see this reflected in the above Eq.~\eqref{eq:UdAV}, where it is manifest that adding any diagonal matrix to $dU_1$ can be compensated for by subtracting the same diagonal matrix from $dV_1^\dagger$. We use this property to fix the diagonal part of $dU_1$ to $0$.

Using the fact that the diagonal part of $dV_1^\dagger$ is purely imaginary and $dS$ is purely real we can deduce:
\begin{align}
    \mathbb{1}\circ\big(S^{-1}U^\dagger dAV-V^\dagger dA^\dagger US^{-1}\big)/2&=\mathbb{1}\circ dV_1^\dagger\\
    \mathbb{1}\circ\big(U^\dagger dAV+V^\dagger dA^\dagger U\big)/2&=dS\label{eq:dS}
\end{align}
where $\circ$ denotes Hadamard product (the element-wise multiplication of matrices). One should be wary  that the object that depends on the SVD being differentiated does not have any dependence on the free gauge. 

Finally, we use the anti-Hermiticity of $dU_1$ and $dV_1$ to find
\begin{align}
    F_{ij}=\frac{1}{S_j^2-S_i^2}\quad \text{for}\quad i\neq j&\quad \text{and}\quad    F_{ii}=0\\
    F\circ\big(SU^\dagger dAV+V^\dagger dA^\dagger US\big)&=dV_1-\mathbb{1}\circ dV_1\\
    F\circ\big(U^\dagger dAVS+SV^\dagger dA^\dagger U\big)&=dU_1\\
\end{align}
The equations for the $dU^\dagger$ and $dV^\dagger$ may be found by simply taking the conjugate of the equations for $dU$ and $dV$.

\section{Review of back-propagation and complex AD}

The basic idea of back-propagation is the efficient use of the chain-rule in differentiating. As an example we take four functions $f(x)$, $x=g(y)$, $y=k(z)$, and $z=p(t)$, and 
consider a function
\begin{equation}
    F(t)=f(g(k(p(t))))\ .
\end{equation}
$x$, $y$, $z$, and $t$ may be real multi-variate variables, but $F$ (and thus also $f$) is assumed to be a single real number. To calculate $F$ at a certain value of $t$ on a computer one would write several function that apply first $p$, then $k$, then $g$, and finally $f$. To determine the derivative of $F$ one would have to deploy the chain rule:
\begin{equation}
    \frac{\partial F(t)}{\partial t_i}=\frac{\partial f(x)}{\partial x_j}\frac{\partial g(y)_j}{\partial y_n}\frac{\partial k(z)_n}{\partial z_m}\frac{\partial p(t)_m}{\partial t_i}\label{eq:chainrule}
\end{equation}
where we assume a sum over repeated indices. This can be efficiently implemented by first calculating the vector $\frac{\partial f(x)}{\partial x_j}$, and consecutively applying to it (on the right) the Jacobian matrices $\frac{\partial g(y)_j}{\partial y_n}$, $\frac{\partial k(z)_n}{\partial z_m}$, and $\frac{\partial p(t)_m}{\partial t_i}$.  
Note that, in principle, the above equation \eqref{eq:chainrule} could be evaluated either from right to left  by first multiplying matrices 
(forward mode AD) or as just described from left to right by applying matrices to the vector, thereby reducing the complexity of the problem
(reverse mode AD, in the opposite direction than evaluation of $F$).

To implement this derivative automatically one needs to know: 
\begin{enumerate}
    \item the order in which the functions were applied,
    \item at which value of the parameters the functions were evaluated,
    \item how the Jacobian 
   $\frac{\partial g(y)_m}{\partial y_i}$  of a function acts on a vector to the left, i.e.\ the map $\Gamma\mapsto\Gamma^m\frac{\partial g(y)_m}{\partial y_i}$.
\end{enumerate}
Points 1.\ and 2.\ are just a bookkeeping problem and easily done automatically in the background. The map from point 3.\ needs to be constructed for any function that is used in constructing $F$. Many functions, like multiplication, addition, or tensor contraction have standard back-propagation built into the available toolboxes. In the subsequent appendices, we will describe specifically how the function for truncated eigendecomposition or truncated SVD are properly back-propagated.

In what follows we will adopt the common convention of calling the accumulated gradient, denoted in Point 3.\ above as $\Gamma$, `adjoint', and labeling it by the variable it is contracted with. For example, if there is a variable $V$ which depends on $X$, we have 
\begin{equation}
    \Gamma_V\frac{\partial V(X)}{\partial X}=\Gamma_X\ .
\end{equation}

In order to incorporate complex numbers into AD, we must take all variables and their complex conjugate as separate degrees of freedom. To see the effect of this we return to Eq.~\eqref{eq:chainrule} and assume that the variables $t_j$ etc.\ include all complex variables and their complex conjugates separately, e.g.\ $t=[a,b,c,\overline{a},\overline{b},\overline{c}]^T$. Eq.~\eqref{eq:chainrule} is thus still the complete derivative. It is then useful to define a linear operator $P$ that flips the index of the variable with its conjugate: $Pt=\overline{t}$ (The dimension of $P$, wherever we use it, is implied by its context). Note that $P^2=\mathbb{1}$. 
We then have that for any arbitrary function $T(t)$ the following holds:
\begin{equation}
    P\frac{\partial T(t)}{\partial t}P = \frac{\partial \overline{T}(t)}{\partial \overline{t}}=\overline{\frac{\partial T(t)}{\partial t}}\ .
\end{equation}

We will assume that we use the AD to differentiate a real-valued cost function $F=f(g(h(k(...))))$, which is made explicit by defining it as $F=(g+\overline{g})/2$, essentially fixing the last function $f$ to be the function that takes the real part of a scalar $x$: $f(x,\overline{x})=(x+\overline{x})/2$.
The adjoints of $f$ can be easily seen to be 
\begin{equation}
    \frac{\partial f(x,\overline{x})}{\partial x}=\frac{\partial f(x,\overline{x})}{\partial \overline{x}}=\frac{1}{2}\ .
\end{equation}
Or, more suggestively, taking $t=[x,\overline{x}]^T$:
\begin{equation}
    \frac{\partial f}{\partial t}P=\overline{\frac{\partial f}{\partial t}}
\end{equation}
If we now consider the subsequent adjoints, $\Gamma_k$ for example, we can see an interesting property under $P$:
\begin{eqnarray}
    \Gamma_kP&=&\frac{\partial f}{\partial g}\frac{\partial g}{\partial h}\frac{\partial h}{\partial k}P\\
    &=&\big(\frac{\partial f}{\partial g}P\big)\big(P\frac{\partial g}{\partial h}P\big)\big(P\frac{\partial h}{\partial k}P\big)\\
    &=&\overline{\frac{\partial f}{\partial g}}\,\overline{\frac{\partial g}{\partial h}}\,\overline{\frac{\partial h}{\partial k}}\\
    &=&\overline{\Gamma_k}
\end{eqnarray}
We thus find that the adjoints of the complex variables are related by complex conjugation to the adjoints of the conjugate variables. Hence, at every back propagation step, the total derivative can be written as certain adjoints and their complex conjugates:
\begin{equation}
    dE = \sum_i \left(\Gamma_{X_i} dX_i + c.c.\right)
\end{equation}
allowing us to only keep track of $\Gamma_{\overline{X}_i}$ (it is convention that $\Gamma_{\overline{X}_i}$ and not $\Gamma_{X_i}$ is stored).

\section{Practical implementation of backwards differentiation of the eigen-decomposition step in Q-deformed CTM}
\label{app:practical_gradient}
In this appendix, we elaborate more on the practical algorithmic implementation of the gradient of the truncated eigendecomposition and ways how to incorporate it within the current AD framework. Using the back-propagation method, our goal is to express the adjoints for the enlarged corner $\overline{M}$, i.e. $\Gamma_{\overline{M}}=\overline{\Gamma_M}$, in terms of adjoints for $C$, $\overline{C}$, $P$, and $\overline{P}$:
\begin{equation}
    \Gamma_{\overline{M}} = \Gamma^C_{\overline{M}} + \Gamma^{\overline{C}}_{\overline{M}} + \Gamma^P_{\overline{M}} + \Gamma^{\overline{P}}_{\overline{M}}
\end{equation}
The contribution from $d\overline{C}$ follows from Eq.~\eqref{eq:dC}:
\begin{equation}
    dC = P^\dagger dM P~\Rightarrow~ \Gamma^{\overline{C}}_{\overline{M}} = P(\Gamma_{\overline{C}})P^\dagger,
\end{equation}
and that same equation tells us $\Gamma^C_{\overline{M}}=0$. Note that due to the $Q$-deformation of CTM, we do not take the diagonal part of $\Gamma_C$ in contrast to what one would usually do for the back-propagation of an eigendecomposition. 

The contribution from $dP$ is less trivial, since it is given by a linear Eq.~\eqref{eq:dP} instead of a closed formula, which we recall here for clarity:
\begin{equation}
    P^\dagger dM^\dagger (\mathbb{1} - PP^\dagger) = C dP^\dagger - dP^\dagger M (\mathbb{1} - PP^\dagger).
\end{equation}
(where we applied a dagger to either side of Eq.~\eqref{eq:dP} and used $C^\dagger=C$ and $M^\dagger=M$)
In fact, there is another equation for $d\overline{P}$ (which must be treated as a separate variable), which also contributes to $\Gamma_M$, obtained by projecting $dM^\dagger$ onto $P^\dagger$ from the right and $(\mathbb{1}-PP^\dagger)$ from the left:
\begin{equation}
    (\mathbb{1} - PP^\dagger) dM^\dagger P =dP C - (\mathbb{1} - PP^\dagger)M dP
\end{equation}
It must be stressed at this point that there is no need to solve the above equation for $dP$ or $d\overline{P}$, as we only really need $\Gamma^{\overline{P}}_{\overline{M}}$ and $\Gamma^{\overline{P}}_{\overline{M}}$. To find $\Gamma^P_{\overline{M}}$ we realize that after vectorization the equation for $d\overline{P}$ becomes:
\begin{eqnarray}
   &&\overrightarrow{dm} =   N \overrightarrow{d\overline{P}} \\
\text{with} &&   dm = \Big[P^\dagger dM^\dagger \left(\mathbb{1}-PP^\dagger\right)\Big]^T \\
\text{and} && N = \left( \mathbb{1}\otimes C-\big[M(\mathbb{1}-PP^\dagger)\big]^T\otimes \mathbb{1} \right)
\end{eqnarray}
and therefore we may rewrite the total derivative as an inner product:
\begin{equation}
    \text{Tr}\left(\Gamma_{\overline{P}} dP^\dagger \right) = \vec{\Gamma_{\overline{P}}}\cdot \vec{d{\overline{P}}} = \vec{\Gamma_{\overline{P}}}\cdot N^{-1} \vec{\overline{dm}}
\end{equation}
and have the matrix $N^{-1}$ act on the left defining a new variable $\vec{\gamma}=N^{-T}\vec{\Gamma_{\overline{P}}} $, where $\vec{\gamma}$ is a vectorized solution to the Sylvester equation of the form:
\begin{equation}
    \Gamma_{\overline{P}} = \gamma C - M(\mathbb{1} - PP^\dagger) \gamma, \label{eq:GammaM}
\end{equation}
In the end we can write that:
\begin{equation}
    \text{Tr}\left(\Gamma_{\overline{P}} dP^\dagger \right) = \text{Tr}\left(\gamma P^\dagger dM^\dagger (\mathbb{1}-PP^\dagger) \right)
\end{equation}
Applying similar tricks to $dP$ we obtain an equation for $\tilde{\gamma}$:
\begin{equation}
    \Gamma_{P} = \tilde{\gamma}C^T - \big[(\mathbb{1} - PP^\dagger)M\big]^T\tilde{\gamma} = \overline{\Gamma_{\overline{P}}}\,,
    \label{eq:gammamtilde}
\end{equation}
remembering that we assume our adjoints to of conjugate variables to be related. We see that the solution of Eq.~\eqref{eq:gammamtilde} can be derived from the solution we got from Eq.~\eqref{eq:GammaM}:
\begin{equation}
    \tilde{\gamma}=\overline{\gamma}
\end{equation}
which together with the contribution from $\overline{P}$ and $\overline{C}$ gives rise to the total $\Gamma_{\overline{M}}$:
\begin{equation}
\begin{split}
     \Gamma_{\overline{M}} = P\Gamma_{\overline{C}}P^\dagger& + (\mathbb{1}-PP^\dagger)\gamma P^\dagger + P \gamma^\dagger (\mathbb{1}-PP^\dagger) \\
     \text{with}\qquad & \Gamma_{\overline{P}} = \gamma C - M(\mathbb{1} - PP^\dagger) \gamma
     \label{eq:gammaMfinal}
\end{split}
\end{equation}

In practice, we can solve the equation for $\gamma$ numerically, without constructing explicitly the matrix $M$, but encoding only its action on $\gamma$. For this we use Matlab's stabilized biconjugate gradients method; since Eq.~\eqref{eq:GammaM} is usually poorly conditioned, 
we use a right preconditioner $C^{-1}$ in order to improve the convergence. Note that constructing the full $\Gamma_{\overline{M}}$, as one would do in fully automated AD, is suboptimal. Clearly $\Gamma_{\overline{M}}$ consists of three terms that are explicitly rank deficient, which could be leveraged to improve the computational complexity of the back-propagation. Note also that the equation previously used~\cite{AD_CTM,Juraj_1AD}, where the truncated spectrum is not taken into account, would consist of only the first term in Eq.~\eqref{eq:GammaM}. 

\section{Practical implementation of backward differentiation of SVD}
Similar to the previous appendix, we present here how to practically implement the formulas for the singular value decomposition. The goal is to take adjoints for $\overline{U}$, $\overline{V}$, and $\overline{S}$: $\Gamma_{\overline{U}}$, $\Gamma_{\overline{V}}$, and $\Gamma_{\overline{S}}$ and map them to adjoints for the original matrix $\overline{A}$ ($\Gamma_{\overline{A}}$).
Taking the previously derived differentials for the SVD, we obtain the following contributions to the adjoints for backwards propagation:
\begin{equation}
  \Gamma_{\overline{A}}=\Gamma_{\overline{A}}^{S}+\Gamma_{\overline{A}}^{U_1}+\Gamma_{\overline{A}}^{V_1}+\Gamma_{\overline{A}}^{diag}+\Gamma_{\overline{A}}^{trunc}  
\end{equation}
The first four terms were also found in \cite{complexSVD}, we present them here again for the readers convenience, and the last term is due to the treatment of truncated spectrum presented in this paper.

We recount:
\begin{eqnarray}
    \Gamma^S_{\overline{A}}&=&\frac{1}{2}U\left(\mathbb{1}\circ(\Gamma_{\overline{S}}+\Gamma_{\overline{S}}^\dagger)\right)V^\dagger\\
    \Gamma^{U_1}_{\overline{A}}&=&U\left(F\circ (U^\dagger\Gamma_{\overline{U}} - \Gamma_{\overline{U}}^\dagger U)\right)SV^\dagger\\
    \Gamma^{V_1}_{\overline{A}}&=&US\left(F\circ (V^\dagger\Gamma_{\overline{V}} - \Gamma_{\overline{V}}^\dagger V)\right)V^\dagger\\ 
    \Gamma^{diag}_{\overline{A}}&=& US^{-1}\left(\mathbb{1}\circ (\Gamma_{\overline{V}}^\dagger V - V^\dagger\Gamma_{\overline{V}})\right) V^\dagger/2
\end{eqnarray}
The last terms follows similarly as Eq.~\eqref{eq:gammaMfinal}:
\begin{equation}
\begin{split}
\Gamma_{\overline{A}}^{trunc} = (\mathbb{1}&-UU^\dagger)\gamma V^\dagger + U\tilde{\gamma}^\dagger(\mathbb{1}-VV^\dagger)\\
  \text{with}\qquad& \Gamma_{\overline{U}}=\gamma S - \underline{A(\mathbb{1}-VV^\dagger)\tilde{\gamma}} \\
  & \Gamma_{\overline{V}} =\tilde{\gamma}S - \underline{A^\dagger(\mathbb{1}-UU^\dagger)\gamma}
\end{split}
\end{equation}
  with the underlined parts added in contrast to previous formulas.

\bibliography{references}
\end{document}